\documentclass[pre,twocolumn]{revtex4}
\usepackage{amsmath}
\usepackage{graphicx}
\newcommand{\ABCD}[4]{\left(\begin{array}{cc}#1&#2\\#3&#4\end{array} \right)}
\makeatletter
\renewenvironment{ruledtabular}{%
 \def\array@default{v}%
 \appdef\tabular@font{\def\@halignto{to\hsize}}%
 \let\tableft@skip@default\tableft@skip
 \let\tableft@skip\tableft@skip@float
 \let\tabmid@skip@default\tabmid@skip
 \let\tabmid@skip\tabmid@skip@float
 \let\tabright@skip@default\tabright@skip
 \let\tabright@skip\tabright@skip@float
 \let\array@row@pre@default\array@row@pre
 \let\array@row@pre\array@row@pre@float
 \let\array@row@pst@default\array@row@pst
 \let\array@row@pst\array@row@pst@float
 \appdef\array@row@rst{%
  \let\array@row@pre\array@row@pre@default
  \let\array@row@pst\array@row@pst@default
  \let\tableft@skip\tableft@skip@default
  \let\tabmid@skip\tabmid@skip@default
  \let\tabright@skip\tabright@skip@default
  \appdef\tabular@font{\let\@halignto\@empty}%
 }%
}{%
}%
\renewcommand\@makefntext[1]{%
  \def\baselinestretch{1}%
  \reset@font\footnotesize
  \parindent 1em%
  \noindent
  \hb@xt@1.8em{\hss\@makefnmark}%
  #1\par
}%
\makeatother
\begin{document}
	\DeclareGraphicsExtensions{.jpg,.pdf,.png,.mps}
\title[Temporal {ABCD} propagation matrices]{Analysis of Optical Pulse Propagation with {ABCD} Matrices}
\author{Shayan Mookherjea}
\email{shayan@caltech.edu}
\homepage{http://www.its.caltech.edu/~shayan}
\affiliation{Department of Electrical Engineering, 136--93 California Institute of Technology, Pasadena, CA~91125}
\author{Amnon Yariv}
\affiliation{Department of Applied Physics, 128--95 California Institute of Technology, Pasadena, CA~91125}
\date{January~3, 2001}
\begin{abstract}
We review and extend the analogies between Gaussian pulse propagation and Gaussian beam diffraction. In addition to the well-known parallels between pulse dispersion in optical fiber and CW beam diffraction in free space, we review temporal lenses as a way to describe nonlinearities in the propagation equations, and then introduce further concepts that permit the description of pulse evolution in more complicated systems. These include the temporal equivalent of a spherical dielectric interface, which is used by way of example to derive design parameters used in a recent dispersion-mapped soliton transmission experiment. Our formalism offers a quick, concise and powerful approach to analyzing a variety of linear and nonlinear pulse propagation phenomena in optical fibers.
\end{abstract}
\maketitle
This paper introduces an \textit{ab-initio\/} study of pulse propagation phenomena analogous to spatial CW diffraction behavior. We address both linear dispersive evolution as well the self-phase modulation effects of the nonlinear index of refraction~\cite{incoll-Crosignani-1990}. The latter is responsible for much of the current interest in nonlinear optical communications, since pulse shapes such as solitons and dispersion-managed solitons display much more attractive transmission properties than linear transmission formats (e.g.~NRZ)~\cite{art-Carter+-1999}.

Such nonlinear pulses are usually self-consistent eigen-solutions of a wave equation, which is the primary reason for their robustness to uncompensated spectral broadening and resultant dissipation into the continuum. The conventional hyperbolic secant soliton is an exact solution of the nonlinear Schr\"odinger equation~\cite{bk-Agrawal}, and propagates indefinitely in a lossless medium without losing its shape. Lossless media can be realized in practice quite effectively by using lumped amplification stages, and erbium-doped fiber amplifiers offer excellent characteristics in this regard.

Breathers, sometimes called dispersion-managed solitons~\cite{art-Suzuki+-1995,art-Smith+-1996}, are also self-consistent `eigen solutions' of the wave equation that propagate with periodic pulse width, chirp etc. While not strictly unchanging in shape, breathers evolve back to their initial configuration, essentially traversing a closed, non-degenerate orbit in phase space~\cite{art-Kutz+-1998}. Unlike pulse shapes designed for linear transmission channels, these pulses do not require periodic dispersion compensation along the transmission channel, and so offer an attractive alternative to the strong control requirements of the nonlinear Schr\"odinger soliton.

Characterizing the solutions of the nonlinear wave equation is often simplest via direct numerical simulation, and this has been particularly true for dispersion mapped solitons~\cite{art-Marcuse-Menyuk-1999}. In order to understand, capture and then predict and utilize the essential physics that guides this behavior, a more conceptually accessible framework is sometimes preferable, such as the variational approach with a pulse shape Ansatz~\cite{art-Anderson-1983}. The pulse shape is described as a dynamical system; we write the Hamiltonian based on the action principle and seek solutions to the Euler-Lagrange equations of motion~\cite{art-Muraki-1991,bk-Goldstein}.
This approach is not always applicable, however, especially when the Ansatz is incapable of capturing some essential physical behavior. Also, it is somewhat more of an analytical tool for probing the dynamics of systems that we already know something about, or can predict at least partially, and it may be convenient to have other approaches that can offer quick insight into constructive aspects of nonlinear propagation, so that different geometries can be analyzed and compared quickly and easily. 

The parallels between dispersive pulse propagation in optical fibers and paraxial CW Gaussian beam diffraction in free space have been identified for some time~\cite{art-Akhmanov+-1968,art-YarivYeh-1978,art-Belanger+-1988}. More recently, the analogies have been extended to include temporal lenses as a way to translate the imaging properties of spatial lenses into the temporal domain~\cite{art-KolnerNazarathy-1989}. In this way, pulse correlation and convolution devices may also be constructed~\cite{art-Lohmann-Mendlovic-1992}. Still more recently, it was shown that temporal lenses can characterize nonlinear effects in the wave equation, leading, for example to the formation of a class of steady-state repeating pulses~\cite{art-Yariv-1999}. We believe that this is perhaps the most potentially useful of the space-time analogies: in this paper, we further extend the use this formalism to describe still more powerful applications such as Gaussian pulse propagation in optical fiber systems, including dispersion mapped systems, including the effects of the nonlinear index of refraction.

We first outline the basic physics that motivates this discussion and sets the context for further development. 

\section{Space-time analogy of beam diffraction and pulse propagation}
\subsection{{CW} Gaussian beam diffraction}
The Fresnel-Kirchoff diffraction integral is a well-founded approach to electromagnetic propagation problems, and several textbooks cover the topic from a variety of approaches~\cite{bk-YarivOE,bk-Haus,bk-Born-Wolf}. We will briefly review only as much as necessary to establish our argument, limiting our argument to diffraction in 1+1 ($x$,$z$) dimensions.

An electromagnetic field of radian frequency $\omega$ and scalar complex amplitude $u(x,z)$ can be represented
\begin{equation}
E(x,z,t) = u(x,z) \, \exp(i \omega t)
\end{equation}
where $u(x)$ obeys the wave equation,
\begin{equation}
\nabla^2 u + k^2 u = 0, \qquad k^2 = \omega^2 \mu \epsilon = \left( \frac{2\pi n}{\lambda}\right)^2.
	\label{eq-wave}
\end{equation}
This equation admits plane wave solutions of the form $\exp(\pm i k z)$ representing propagation along $\mp z$ respectively, and indeed, an arbitrary superposition of plane waves, each with the same wavelength, propagating along all possible directions,
\begin{equation}
u(x,z) = \int \tilde{u}_0(k_x) \exp[i(k_x x )- i \sqrt{k^2-k_x^2} z] \, dk_x
	\label{eq-outfield-infield}
\end{equation}
where $\tilde{u}_0$ is the Fourier transform of the input field $u_0(x,0)$. 

We consider optical beams whose plane wave components propagate at small angles to the $z$ axis (paraxial approximation), so that we can expand the square root in~(\ref{eq-outfield-infield}) in a Taylor series and keep the first two terms, 
\begin{eqnarray}
E(x,z)&\\
=& e^{i \omega t -i k z} \int \left[\tilde{u}_0(k_x) \exp\left(i \frac{k_x^2}{2k} z \right) \right] \exp(ik_x x) \, dk_x \nonumber \\
=& e^{i\omega t -i k z} \left[\sqrt{\frac{i k}{2 \pi z}} \int u_0(x') \exp\left[ -\frac{i k (x-x')^2}{2 z}  \right] \, dx' \right] \nonumber
\end{eqnarray}
where the term in parentheses defines $u(z,t)$, the field envelope,
\begin{equation}
u(z,t)=\sqrt{\frac{i k}{2 \pi z}} \int u_0(x') \exp\left[ -\frac{i k}{2 z} (x-x')^2 \right] \, dx'
\label{eq-define-diffract}
\end{equation}

The propagation of continuous-wave (CW) Gaussian beams in free space and rotationally-symmetric quadratic graded-index media is conveniently described by assuming that the envelope has the form~\cite{bk-YarivOE}
\begin{equation}
u = \exp\left\{ -i \left[ P(z) + \frac{k}{2 q(z)} r^2\right] \right\}
\end{equation}
where, we find by substitution into the wave equation~(\ref{eq-wave}) that $dP/dz = -i/q(z)$ for such media. The $q$-parameter describes the Gaussian beam completely,
\begin{equation}
\frac{1}{q(z)} = \frac{1}{R(z)} - i \frac{\lambda}{\pi n w^2(z)}.
	\label{eq-def-q-spatial}
\end{equation}
In the above definition, $R(z)$ describes the radius of curvature of the beam, and $w(z)$ the beam spot size. 

The usefulness of the $q$-parameter lies in the bilinear transformation (ABCD law) that characterizes how this parameter evolves with propagation. For an optical system described by a real (or complex) ABCD matrix, the output $q$ parameter is given by
\begin{equation}
q_{o} = \frac{A q_i+B}{C q_i + D}.
\end{equation}
Separating the real and imaginary parts of $q_o$ enables us to calculate the radius of curvature and spot size of the Gaussian beam at the output of the optical system. Many practically important optical systems and their corresponding phenomena can be described by simple ABCD matrices, such as propagation in a uniform medium, focusing via a thin lens, beam transformation at a dielectric interface, propagation through a curved dielectric interface and thick lens, propagation in a medium with a quadratic index variation etc.~\cite[Table~2-1]{bk-YarivOE}

\subsection{Gaussian pulse propagation}
Consider a single mode in an optical fiber, usually the lowest-order fundamental mode, excited at $z=0$, and with an assumed temporal envelope of the Gaussian form,
\begin{equation}
E(z=0,t) = \mathbf{Re}\left[\exp(-\alpha t^2 + i \omega_0 t) \right]
\end{equation}
and write as a Fourier transform integral,
\begin{equation}
E(0,t) = \mathbf{Re}\left[\exp(i \omega_0 t) \int \tilde{u}_0(\Omega) \exp(i \Omega t)\, d\Omega\right]
\end{equation}
where $\tilde{u}_0$ is the Fourier transform of the Gaussian envelope $u_0=\exp(-\alpha t^2)$.

As in the spatial case, we can be think of this as a superposition of time-harmonic fields, each with frequency $(\omega_0+\Omega)$ and amplitude $\tilde{u}_0(\Omega) d\Omega$. These waves will experience a phase delay when propagating a distance $z$; we multiply each frequency component by its propagation delay factor $\exp[-i \beta(\omega_0+\Omega) z]$ so that
\begin{equation}
E(z,t) = \int \tilde{u}_0(\Omega) \exp[i(\omega_0+\Omega) t -i \beta(\omega_0 + \Omega) z]\, d\Omega.
\end{equation}

Expanding $\beta(\omega_0+\Omega)$ in a Taylor series about the optical frequency $\omega_0$,
\begin{equation}
\beta(\omega_0+\Omega) = \beta(\omega_0) + \left.\frac{d\beta}{d\omega}\right|_{\omega=\omega_0} \Omega + \frac{1}{2} \left. \frac{d^2\beta}{d\omega^2}\right|_{\omega=\omega_0} \Omega^2 + \ldots
	\label{eq-beta-Taylor}
\end{equation}
we can write
\begin{eqnarray}
E(z,t) &
=& \exp[i(\omega_0 t - \beta_0 z)] \times \label{eq-E-solution} \\
&&\int \tilde{u}_0(\Omega) \exp\left\{ i \left[\Omega t - \beta' \Omega z - \frac{1}{2} \beta'' \Omega^2 z\right]\right\} \, d\Omega \nonumber
\end{eqnarray}
where the integral defines the field envelope $u(z,t)$, so that
\begin{equation}
E(z,t) = u(z,t)\, \exp[i(\omega_0 t- \beta_0 z)].
\end{equation}
The differential equation satisifed by $u$ is, to second order of derivatives of $\beta$~\cite{bk-Agrawal},
\begin{equation}
\frac{\partial u}{\partial z}+\beta' \frac{\partial u}{\partial t}+\frac{1}{2}\beta'' \frac{\partial^2 u}{\partial t^2}=0.
	\label{eq-propagation-equation}
\end{equation}
The solution~(\ref{eq-E-solution}) can be written  using the inverse Fourier transform relationship,
\begin{eqnarray}
u(z,t) &=& \int \left[ \tilde{u}_0(\Omega) \exp\left( -\frac{i}{2} \beta'' \Omega^2 z- i \beta' \Omega z\right) \right] 
\nonumber \\
&&\qquad \times \exp(i \Omega t) \, d\Omega \nonumber \\
&=& \displaystyle \sqrt{\frac{1}{i 2 \pi \beta'' z}} \nonumber \\
&&\quad \times \int u_0(t') \exp\left[ \frac{i}{2\beta'' z}(T-t')^2\right] \, dt'
	\label{eq-define-disperse}
\end{eqnarray}
where $T=t-\beta' z = t-z/v_g$ is the time coordinate in the frame of reference co-moving with the pulse envelope at the group velocity $v_g = 1/\beta'$. Dispersion of the group-velocity (GVD) is represented by $\beta''$.

The formal similarity between~(\ref{eq-define-diffract}) and~(\ref{eq-define-disperse}) is the principal motivation for this analysis. We can write down a set of space-time translation rules {(see Table~\ref{tab-space-time-rules})} to apply results from spatial diffraction to temporal dispersion and vice versa. 
One family of results that can be derived from this space-time analogy correspond to spatial imaging e.g.~the 2-$f$ and 4-$f$ optical systems. These can be applied to pulse compression or expansion experiments etc~\cite{art-KolnerNazarathy-1989}. 
 
\begin{table*}
\caption{Space-time translation rules}
	\label{tab-space-time-rules}
\begin{ruledtabular}
\begin{tabular}{|cc|cc|}
spatial frequency (Fourier variable)	& $k_x$ & $\Omega$	& frequency (Fourier variable) \\
transverse distance	& $x$		& $t-\frac{z}{v_g}$	& time (in moving reference frame) \\
propagation distance	& $z$		& $z$			& propagation distance \\
wavevector (inverse)	& $k^{-1}$	& $-\beta''$		& GVD coefficient (negative)
\end{tabular}
\end{ruledtabular}
\end{table*}

But we will see in later sections that many linear and nonlinear pulse propagation systems can be described by cascading simple ABCD matrices, and this can result in substantially simpler calculations and more direct physical understanding of the physical processes involved in nonlinear pulse propagation. We will first need to develop some additional facility in characterizing optical systems associated with the pulse propagation equations.

The spatial $q$-parameter has a temporal equivalent $q_t$ in accordance with the space-time translation rules of Table~\ref{tab-space-time-rules}, defined by
\begin{equation}
\frac{1}{q_t(z)} = \frac{1}{R_t(z)} + i\frac{2 \beta''}{\tau^2(z)},
	\label{eq-def-q-temporal}
\end{equation}
where $\tau(z)$ represents the pulse width (scaled in the $T$ frame by $\sqrt{2}$) and $R_t(z)$ its chirp. A Gaussian pulse in linearly dispersive fibers is then represented by the envelope~\cite{art-Yariv-1999}
\begin{eqnarray}
u(z,t) &=&u_0 \frac{\tau_0}{\tau(z)}\\
&& \exp\left[ i \tan^{-1} \frac{z}{\zeta_0} + i \frac{t^2}{2 \beta'' R_t(z)}+ \frac{\beta''}{|\beta''|} \frac{t^2}{\tau^2(z)}\right].
\nonumber
\end{eqnarray}
where the pulse width and chirp satisfy evolution equations in linear dispersive fibers exactly analogous to their spatial counterparts, beam spot size and radius of curvature, in free space~\cite{art-Yariv-1999}
\begin{eqnarray}
\tau^2(z) &=& \tau_0^2\left( 1+ \frac{z^2}{\zeta_0^2}\right), \nonumber \\
R_t(z) &=& z\left( 1+\frac{\zeta_0^2}{z^2}\right)
	\label{eq-width-chirp-propagate}
\end{eqnarray}
with $\zeta_0=\tau_0^2/2|\beta''|$ defining the dispersion length~\cite{bk-Agrawal}. 

\section{Components of the ABCD formalism for Gaussian pulse propagation}
As a simple example of the application of the above translation rules, we consider the propagation of a Gaussian input pulse with envelope
\begin{equation}
U(0, T) = \exp \left( -\frac{T^2}{2 T_0^2} \right).
\end{equation}
The transmission medium comprises of two concatenated sections of fiber with lengths $z_1$ and $z_2$ and with GVD coefficients $\beta_1''$ and $\beta_2''$ respectively. We ignore any nonlinear effects in this simple problem, and assume that the medium is lossless. What are the pulse characteristics at the output of the second medium i.e.~what is the pulse width at $z=z_1+z_2$?

One way of solving this problem is by recourse to the wave equation~(\ref{eq-define-disperse}) solution by the Fourier transform technique. We have,
\begin{eqnarray}
\tilde{U}(z_2, \omega) &=& \tilde{U}(z_1,\omega) \exp \left( \frac{i}{2} \beta_2'' z_1 \omega^2 \right) \nonumber \\
&=& \tilde{U}(z_0, \omega) \exp \left[ \frac{i}{2} \left( \beta_1'' z_1 + \beta_2'' z_2 \right) \omega^2 \right].
\end{eqnarray}

Taking the inverse Fourier transform,
\begin{equation}
\begin{array}{lcl}
U(z_1+z_2, T) &=& \\
\multicolumn{3}{l}{\displaystyle \frac{1}{2\pi} \int_{-\infty}^{\infty} \tilde{U}(0, \omega) 
\left[ \frac{i}{2} \left( \beta_1'' z_1 + \beta_2'' z_2 \right) \omega^2 -i \omega T\right] \, d\omega}\\
&=& \displaystyle {\left[\frac{T_0^2}{T_0^2-i(\beta_1'' z_1 + \beta_2'' z_2)}\right]}^{\frac{1}{2}} \\
&&  \displaystyle \times
\exp \left[ - \frac{T^2}{2(T_0^2 - i(\beta_1'' z_1+\beta_2'' z_2)}\right],
\end{array}
\end{equation}
from which we see that the ratio of the output to input pulse width, therefore, is
\begin{equation}
\frac{T_1}{T_0} = {\left[ 1+ \left( \frac{\beta_1'' z_1 + \beta_2'' z_2}{T_0^2}\right)^2\right]}^{\frac{1}{2}}.
	\label{eq-width-ratio-fourier}
\end{equation}

We will now verify~(\ref{eq-width-ratio-fourier}) using the {ABCD} matrix approach. The system is described very simply by the product of three matrices,
\begin{eqnarray}
M &=& \ABCD{1}{z_2}{0}{1}.\ABCD{1}{0}{0}{\frac{\beta_2''}{\beta_1''}}.\ABCD{1}{z_1}{0}{1} \nonumber \\
&=& \ABCD{1}{z_1+\frac{\beta_2''}{\beta_1''}z_2}{0}{\frac{\beta_2''}{\beta_1''}}
\end{eqnarray}
so that
\begin{eqnarray}
q_2 = \frac{A q_1+B}{C q_1+D} = \frac{\beta_1''}{\beta_2''}q_1 + \frac{\beta_1''}{\beta_2''} z_1 + z_2.
\end{eqnarray}

Using the shorthand notation
\begin{equation}
\begin{array}{cc}
R_2 \equiv R(z_1+z_2), & R_1 \equiv R(0), \\
\tau_2 \equiv \tau(z_1+z_2), & \tau_1 \equiv \tau(0).
\end{array}
\end{equation}
we have
\begin{equation}
\frac{\frac{1}{R_2} - i \frac{2\beta_2''}{\tau_2^2}}{\left(\frac{1}{R_2}\right)^2+\left(\frac{2\beta_2''}{\tau_2^2}\right)^2}
= \frac{\beta_1''}{\beta_2''}
\frac{\frac{1}{R_1} - i \frac{2\beta_1''}{\tau_1^2}}{\left(\frac{1}{R_1}\right)^2+\left(\frac{2\beta_1''}{\tau_1^2}\right)^2}
+\frac{\beta_1''}{\beta_2''}z_1+z_2
	\label{eq-disp-real-imag}
\end{equation}

The real and imaginary parts of both sides of the above equation have to be equal, leading to a pair of simultaneous equations. For an unchirped input pulse, $R_1=0$ so that equality of the imaginary parts leads to
\begin{equation}
\left(\frac{1}{R_2}\right)^2+\left(\frac{2\beta_2''}{\tau_2^2}\right)^2 = \left( \frac{2\beta_2''}{\tau_2\tau_1}\right)^2.
	\nonumber
\end{equation}
Subsituting this expression into the equation of equality of the real parts of~(\ref{eq-disp-real-imag}) and some algebraic manipulation leads to
\begin{equation}
\frac{\tau(z_1+z_2)}{\tau(0)}=
\frac{\tau_2}{\tau_1}=\left[ 1+ {\left(\frac{\beta_1'' z_1+\beta_2'' z_2}{\tau_1^2/2}\right)}^2\right]^{\frac{1}{2}}
\end{equation}
which is the same as~(\ref{eq-width-ratio-fourier}), since $\tau = \sqrt{2}\, \Delta T$.

In the above calculation, we have carried out some algebraic simplifications by hand in order to show that the result obtained by the ABCD matrix approach is the same as that obtained by the Fourier transform approach. Nevertheless, the former is computationally much simpler, and separating the real and imaginary parts of~(\ref{eq-disp-real-imag}) as part of a numerical algorithm can be carried out without the notational complexity of, for example, rationalizing the denominator. 

While second-order dispersion is conveniently represented by the ABCD matrix approach, there are problems with extending the analysis to higher orders of dispersion. The slowly-varying envelope equation analogous to~(\ref{eq-propagation-equation}) including the effects of third-order dispersion
\begin{equation}
i \frac{\partial U}{\partial z} = \frac{1}{2}\beta'' \frac{\partial^2 U}{\partial T^2}+\frac{i}{6}\beta'''\frac{\partial^3 U}{\partial T^3}
\end{equation}
or its solution in terms of the Fourier transformed variables,
\begin{equation}
\tilde{U}(z, \omega) = \tilde{U}(0, \omega) \exp\left(\frac{i}{2}\beta'' z \omega^2 +\frac{i}{6}\beta''' z \omega^3\right)
	\label{eq-tod-fourier}
\end{equation}
does not have an equivalent in the CW spatial diffraction context.

To see this, consider the next term in the Taylor expansion of $\sqrt{k^2 - k_x^2}$ in~(\ref{eq-outfield-infield}), which leads to an expression of the form
\begin{eqnarray}
u_2(x) &=& e^{-i k z} \times \\
&& \int u_1(k_x) \exp\left(i \frac{k_x^2}{2 k}z+ i \frac{k_x^4}{8 k^3}z\right)\, \exp(i k_x x) \, dk_x. \nonumber
\end{eqnarray}
Using the space-time translation rules, we find that the above expression contains a description of second and {\textit{fourth}}-order dispersion, not third-order dispersion.

This is obviously a general characteristic of the above Taylor expansion;  all odd-order dispersion terms have no spatial paraxial diffraction equivalent in the ABCD matrix content. Recall that the effect of $\beta'$ is accounted for by transforming to a moving reference frame $T=t-\beta' z$. 

For completeness, we derive the translation rule for {\textit{any}} even-order dispersion in terms of the equivalent term in CW diffractive optics. A little algebra will show that the generalization of~(\ref{eq-tod-fourier}) yields
\begin{eqnarray}
\tilde{U}(z, \omega) &=& \tilde{U}(0, \omega) \times 	\label{eq-high-od-fourier}\\
&&\exp\left(\frac{i}{2}\beta''  \omega^2 z+\frac{i}{6}\beta''' \omega^3 z \right. \nonumber \\
&&\quad \left.+\ldots+\frac{i}{(2r)!}\beta^{(2r)} \omega^{2r}z\right),
\end{eqnarray}
and, correspondingly, for the diffraction of a Gaussian beam,
\begin{eqnarray}
\tilde{U}(x, k_x) &=& \tilde{U}(0, k_x)
\exp\left[i \frac{k_x^2}{2 k}z+ i \frac{k_x^4}{8 k^3}z +\ldots \right. \nonumber \\
&&\quad \left.+ i \frac{(-1)^r}{r!}\prod_{l=0}^{r-1}\left(\frac{1}{2}-l\right) \frac{k_x^{2r}}{k^{2r-1}}z\right].
\end{eqnarray}
Therefore, the translation rule for $2r$-order temporal dispersion is given by
\begin{eqnarray}
\beta^{(2r)} &\leftarrow & \left[ (-1)^r \frac{(2r)!}{r!} \prod_{l=0}^{r-1}\left(\frac{1}{2}-l\right) \right] \frac{1}{k^{2r-1}}
	\nonumber \\
&=& - \left[ \frac{(2r)!}{r!\, 2^r} \prod_{l=1}^{r-1}(2l-1) \right] \frac{1}{k^{2r-1}}.
\end{eqnarray}

Note that this corrects the statement in~\cite{art-KolnerNazarathy-1989}:
\begin{quote}
The slowly varying envelope equations corresponding to modulated plane waves in dispersive media have the same form as the paraxial equations describing the propagation of monochromatic waves of finite spatial extent (diffraction).
\end{quote}
We append that this correspondence holds for all even orders of dispersion, and of course, for $\beta'$ as well, by transforming to a moving reference frame.

Our ABCD formalism would be of limited interest if the only phenomena it could capture were that of dispersive propgation. But, as mentioned in an earlier section, the development of the time-lens formalism lets us describe nonlinear mechanisms as well.

By analogy to spatial lenses which are characterized by a lens factor $\exp(i k r^2/2f)$ which multiplies an incoming optical beam, we define a temporal lens as a device that multiplies the pulse envelope by a factor~\cite{art-Yariv-1999,art-KolnerNazarathy-1989}
\begin{equation}
{\mathrm{Lens~Factor}} = \exp\left[ - i \frac{t^2}{2\beta'' f_t} \right] \equiv \exp\left[-i b t^2\right]
\end{equation}

The ABCD matrix representing a temporal lens has the same form as that of a spatial lens,
\begin{equation}
M=\displaystyle \ABCD{1}{0}{-\frac{1}{f_t}}{0}
\end{equation}
where $f_t$ represents the temporal ``focal length''. 
\begin{figure}
\begin{center}
\scalebox{1}{\resizebox{\linewidth}{!}
 {\includegraphics{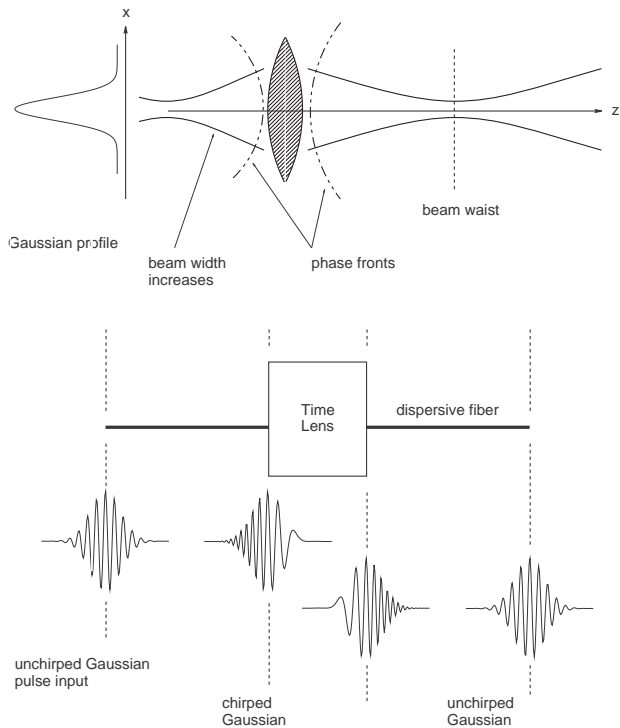}}} 
\caption{(a) Spatial lens (b) Temporal lens} 
	\label{fig-lens}
\end{center}
\end{figure}

A comparison of spatial and temporal lensing is shown in Figure~\ref{fig-lens}.  In the spatial case, the lens compensates for the spreading of the beam waist, and ``flips'' the phase fronts to convert a diverging beam into a converging one. Similarly, a temporal lens reverses the sign of the chirp, so that further propagation in a $\beta''<0$ dispersive fiber will compensate for the chirp (phase modulation) caused this far. This is also an interesting and physically illuminating approach to discussing the physics of the formation of solitons~\cite[Chapter~19]{bk-YarivOE}.

One possible implementation, as proposed in~\cite{art-Yariv-1999}, is to achieve temporal lensing by self-phase modulation during the passage of the pulse through a section of nonlinear fiber ($\beta''\approx 0$, $n_2>0$). For short distances, $z\ll \pi \tau_0^2/|\beta''|$ and when $\beta''/\tau_0^2\ll (2\pi n_2/\lambda) I_p$ for peak intensity $I_p$, a pulse with input electric field envelope $u(0, T)$ emerges from a length $z$ of nonlinear fiber with phase modulation
\begin{equation}
u(z, T) = u(0, T) \exp\left[ -i \frac{\omega_0 n_2 z}{2 c \eta} {\left| u(0, T)\right|}^2 \right]
	\label{eq-define-SPM}
\end{equation}
where $\eta= \sqrt{\mu/\epsilon}$ defines the impedance of free space. 
If we write the pulse intensity as
\begin{equation}
I=\frac{|u|^2}{2\eta}=I_p \exp\left[ -2 \left(\frac{T}{\tau_0}\right)^2\right]
\end{equation}
and keep the first two terms in the Taylor expansion of the exponential in~(\ref{eq-define-SPM}), 
\begin{equation}
u(z,T)=u(0,T) \exp\left[ i \frac{2 \omega_0 n_2 I_p z}{c \tau_0^2}T^2\right]
\end{equation}
modulo a phase term linear in $z$ that is independent of $T$. The effect of propagation through length $L$ of nonlinear fiber is to impart a quadratic chirp to the pulse, which we represent by the multiplicative term $\exp(-i b t^2)$ so that
\begin{equation}
b = \frac{1}{2 \beta'' f_t}=-\frac{2 \omega_0 n_2 I_p L}{c \tau_0^2}.
	\label{eq-define-b}
\end{equation}

Another method of obtaining time lensing is based on the principle of electro-optic modulation~\cite{art-KolnerNazarathy-1989}. An electro-optic phase modulator driven by a sinusoidal bias voltage of angular frequency $\omega_m$ results in a phase modulation that is approximately quadratic under either extremum of the sinusoid. The phase shift can be written as
\begin{equation}
\exp[i \phi(t)] = \exp\left[-i K \left( 1-\frac{\omega_m^2 t^2}{2}\right)\right]
\end{equation}
where $K$ is the modulation index~\cite[\S9.4]{bk-YarivOE}. In this case,
\begin{equation}
b= \frac{2}{K \omega_m^2}
\end{equation}

We have described our temporal lens by a section of nonlinear fiber of $\beta'' \approx 0$, analogous to a spatial thin lens, which is assumed to have no thickness. Just as practical lenses do have some thickness, practical fibers have non-zero $\beta''$. For those situations in which this cannot be ignored, or may even be utilized constructively, we derive the corresponding equivalent of a spatial ``thick lens''. 

Our first step is to characterize the temporal equivalent of a curved dielectric interface: a spatial lens comprises of two such interfaces separated by a length of material of enhanced refractive index. At a planar dielectric interface between two media of refractive indices $n_1$ and $n_2$, a Gaussian beam undergoes a change in the radius of curvature, but is unchanged in beam width,
\begin{equation}
R_2 = \frac{n_2}{n_1} R_1, \qquad w_1=w_2.
\end{equation}
By analogy, a chirped Gaussian pulse at the interface between two fibers of GVD coefficients $\beta_1''$ and $\beta_2''$ transforms to a different chirp, but with unchanged pulse width,
\begin{equation}
\frac{1}{\beta_2'' R_2} = \frac{1}{\beta_1'' R_1}, \qquad \tau_1 = \tau_2.
\end{equation}
Of course, the pulse width evolves differently in the two sections of fiber,
\begin{equation}
\tau_i^2(z) = \tau_{0i}^2\left( 1 + \frac{z^2}{\zeta_{0i}^2} \right) \qquad {i=1,2}
\end{equation}
where $\zeta_{0i}$ is the dispersion length in fiber $i$. 

The ABCD matrix for a (spatial) spherical dielectric interface and its temporal translation are
\begin{equation}
M: \quad \ABCD{1}{0}{\frac{n_2-n_1}{n_2 R}}{\frac{n_1}{n_2}} \mapsto \ABCD{1}{0}{\frac{1-\beta_2''/\beta_1''}{R_l}}{\frac{\beta_2''}{\beta_1''}}
\end{equation}
What does this represent? We use the ABCD bilinear transformation,
\begin{equation}
q_2 =q_1\left/ \left[\frac{1}{R_l}\left( 1-\frac{\beta_2''}{\beta_1''}\right) q_1 + \frac{\beta_2''}{\beta_1''}\right] \right.
\end{equation}
which implies that
\begin{eqnarray}
\frac{1}{q_2} &=& \left( \frac{1}{R_2} + i \frac{2\beta_2''}{\tau_2^2}\right) \nonumber \\
&=&\frac{1}{R_l}\left(1 - \frac{\beta_2''}{\beta_1''}\right) + \left(\frac{1}{R_1}+i \frac{2\beta_1''}{\tau_1^2}\right) \frac{\beta_2''}{\beta_1''}
\end{eqnarray} 
After some algebraic manipulation, we can write the above as
\begin{equation}
\beta_1''\left( \frac{1}{R_l} - \frac{1}{R_2}\right) = \beta_2'' \left(\frac{1}{R_l}-\frac{1}{R_1} \right)
\end{equation}
showing explicitly how the chirp transforms at this interface. 

The ABCD matrix for a temporal lens of ``thickness'' $d$ is written as the product of three ABCD matrices
representing, when read from right to left, a transition from the input fiber to the fiber that defines the thin temporal lens, propagation in the second fiber, and a transition back to the input fiber,
\begin{eqnarray}
M &=& \nonumber \\
&&\ABCD{1}{0}{\frac{1-\beta_1''/\beta_2''}{R_2}}{\frac{\beta_1''}{\beta_2''}}.\ABCD{1}{d}{0}{1}.
\ABCD{1}{0}{\frac{1-\beta_2''/\beta_1''}{R_1}}{\frac{\beta_2''}{\beta_1''}} \nonumber
\end{eqnarray}
and, multiplying the matrices together, we get a single ABCD matrix which defines the output $q_t$ parameter via the usual bilinear transformation $(A q_t + B) / (C q_t + D)$,
\begin{widetext}
\begin{equation}
M=\ABCD{1+\frac{d}{R_1}\left(1-\frac{\beta_2''}{\beta_1''}\right)}
{d \frac{\beta_2''}{\beta_1''}}
{\frac{d}{R_1(-R_2)}\left(\frac{\beta_2''}{\beta_1''}-1\right) \left(1-\frac{\beta_1''}{\beta_2''}\right) - 
\left(\frac{1}{R_1}+\frac{1}{-R_2}\right) \left(1-\frac{\beta_1''}{\beta_2''}\right)}
{1+\frac{d}{-R_2}\left(1-\frac{\beta_2''}{\beta_1''}\right)}
\end{equation}
The temporal focal length $\hat{f_t}$ is analogous to the spatial focal length and is given by $-A/C$,
\begin{equation}
\hat{f_t} = \left[1+\frac{d}{R_1}\left(1-\frac{\beta_2''}{\beta_1''}\right)\right] \left/ \left[
\left(\frac{1}{R_1}+\frac{1}{-R_2}\right) \left(1-\frac{\beta_1''}{\beta_2''}\right)-
\frac{d}{R_1(-R_2)}\left(\frac{\beta_2''}{\beta_1''}-1\right) \left(1-\frac{\beta_1''}{\beta_2''}\right)
\right]\right.
\end{equation}
\end{widetext}
The temporal focal length defines the time from the output plane at which an initially unchirped pulse becomes unchirped again. 

We can write the above in slightly simpler notation, for the specific case $R_1=-R_2=R$, and let $\kappa = d/R$, $\Delta\beta''=\beta_2''-\beta_1''$,
\begin{equation}
\hat{f}_t = \left[ \frac{\displaystyle 1-\kappa \frac{\Delta\beta''}{\beta_1''}}{\displaystyle 1-\frac{\kappa}{2}\frac{\Delta\beta''}{\beta_1''}} \frac{\beta_2''}{\Delta\beta''}\right] \frac{R}{2}
\end{equation}
where the term in parentheses represents an enhancement factor over the ``thin lens'' formula.

For $\kappa \ll 1$, we can simplify the above expression keeping terms of $O(\kappa)$,
\begin{eqnarray}
\frac{1}{f_t} &\approx& \left( 1- \frac{\kappa}{2} \frac{\Delta \beta''}{\beta_1''}\right)
\left( 1+ \kappa\frac{\Delta \beta''}{\beta_1''}\right) \frac{2 \Delta \beta''}{R\, \beta_2''} \nonumber \\
&\approx& \left( 1+ \frac{\kappa}{2} \frac{\Delta \beta''}{\beta_1''}\right) \frac{2 \Delta \beta''}{R\, \beta_2''}
\end{eqnarray}
The above relation confirms our physical intuition that if $\beta_2''-\beta_1''=\Delta\beta'' <0$, then we have reduced $f_t$, the distance to the point of zero chirp from the output plane, for an initially unchirped input pulse. 

We now have the tools we need to analyze a reasonably complicated practical problem: designing the length of a dispersion map so as to get self-consistent eigen-pulses with periodic pulse width and chirp. 

\section{Dispersion-managed soliton transmission experiment}
It has been recently found that a stable, self-consistent pulse solution exists in a dispersion-managed fiber transmission system~\cite{art-Smith+-1996}. While these are not solitons in the strict mathematical sense, they have been called dispersion-managed solitons, or perhaps more appropriately, breathers. They demonstrate periodic behaviour: the pulse width and chirp of Gaussian breathers, for instance, are periodic functions of the propagation distance. Breathers share a property in common with solitons in that they can propagate indefinitely without losing shape; even though the pulse shape undergoes changes within a disperion map period, the pulse does not disperse away to infinity, or tend to self-focus to a point either of which invalidate the applicability of the nonlinear Schr\"odinger equation after a certain distance.

A dispersion-mapped (DM) soliton is closer to a Gaussian shape than the hyperbolic secant of the nonlinear Schr\"odinger equation~\cite{art-Lakoba-Kaup-1998}, and it is interesting to ask whether our analysis is capable of capturing the essential aspects of its evolution along a dispersion-mapped transmission channel.

We consider, as our example, the paper by Mu et al.~\cite{art-Mu+-2000} who have simulated DM soliton dynamics in a recirculating fiber loop. Their dispersion map consists of 100~km of dispersion shifted fiber (SMF-LS) with normal dispersion $D_1$ equal to -1.10~ps/nm-km at 1551~nm, followed by an ``approximately 7-km span'' of standard single-mode fiber (SMF-28) with an anomalous dispersion $D_2$ equal to 16.6~ps/nm-km at 1551~nm. The results of the paper indicate that Gaussian shaped pulses of pulse duration 5.67~ps and peak power 9~dBm were used. We will derive the result that, for these parameters and given the length of SMF-LS fiber, the length of SMF-28 fiber that needs to be used is indeed ``approximately 7-km''.  In other words, we will show that this given dispersion map can support lowest-order chirped Gaussian self-consistent solutions, i.e.~breathers.

\begin{figure}
\begin{center}
\scalebox{1}{\resizebox{\linewidth}{!}
 {\includegraphics{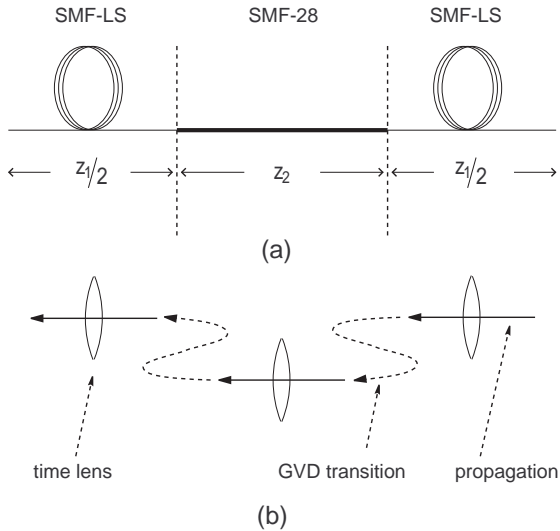}}} 
\caption{(a) Analytical schematic of dispersion map from~~\cite{art-Mu+-2000} and (b) its representation
to express in terms of ABCD matrix elements.} 
	\label{fig-dispersion-map}
\end{center}
\end{figure}

The dispersion map, shown schematically in Figure~\ref{fig-dispersion-map}, consists of three fiber segments: a length $z_1/2$ equal to 50~km of SMF-LS fiber, followed by a length $z_2$ of SMF-28 fiber, whose numerical value is to be determined, and then the remainder $z_1/2$ of SMF-LS fiber. Each segment of fiber has nonlinear characteristics, which we model via a time lens situated, for simplicity at the individual midpoints of the respective segments. Consequently, each segment is described by the cascaded product of three ABCD matrices, with two additional matrices representing the transitions between fibers of different $\beta''$. For simplicity, we will assume that the nonlinear properties of the fibers are identical.

The overall ABCD matrix for the system can be written down quite easily,
\begin{eqnarray}
M &=& \ABCD{1}{z_1/4}{0}{1} . \ABCD{1}{0}{-1/f_t}{1} . \ABCD{1}{z_1/4}{0}{1}  \nonumber \\
&&\quad .\ABCD{1}{0}{0}{\beta_1''/\beta_2''}.\ABCD{1}{z_2/2}{0}{1}.\ABCD{1}{0}{-1/f_t}{1}\nonumber \\
&&\quad .\ABCD{1}{z_2/2}{0}{1}.\ABCD{1}{0}{0}{\beta_2''/\beta_1''}
.\ABCD{1}{z_1/4}{0}{1} \nonumber \\
&&\qquad . \ABCD{1}{0}{-1/f_t}{1} . \ABCD{1}{z_1/4}{0}{1}
\end{eqnarray}
which, after some algebra, can be written as an ABCD matrix with the following elements,
\begin{widetext}
\begin{eqnarray}
A &=& \left[ \left(1-\frac{z_1}{4 f}\right) \left(1-\frac{z_2}{2 f}\right) - \frac{z_1}{4 f}\frac{\beta_1''}{\beta_2''} \left(2-\frac{z_1}{4 f}\right) \right] \left(1-\frac{z_1}{4 f}\right) \nonumber \\
&& \quad -\left[ \frac{\beta_2''}{\beta_1''}\left(1-\frac{z_1}{4 f}\right)\left(2-\frac{z_2}{2 f}\right)\frac{z_2}{2 f}
+ \left(2-\frac{z_1}{4 f}\right) \left(1-\frac{z_2}{2 f}\right) \frac{z_1}{4 f}\right] \label{eq-A-expr} \\
D &=& -\frac{z_1}{4 f} \left[ \left(1-\frac{z_2}{2 f}\right) + \frac{\beta_1''}{\beta_2''} \left(1-\frac{z_1}{4 f}\right) \right] \left(2-\frac{z_1}{4 f}\right) 
- \left[ \frac{\beta_2''}{\beta_1''} \frac{z_2}{2 f} \left(2-\frac{z_2}{2 f}\right) - \left(1-\frac{z_2}{2 f}\right) \left(1-\frac{z_1}{4 f}\right) \right] \left(1-\frac{z_1}{4 f}\right) \label{eq-D-expr}
\end{eqnarray}
\begin{eqnarray}
B &=& \left[\left(1-\frac{z_1}{4 f}\right) \left(1-\frac{z_2}{2 f}\right) - \frac{z_1}{4 f} \frac{\beta_1''}{\beta_2''} \left(2-\frac{z_1}{4 f}\right) \right] \frac{z_1}{4} \left(2-\frac{z_1}{4 f}\right) \nonumber \\
&&\quad +\left[ \frac{\beta_2''}{\beta_1''} \left(1-\frac{z_1}{4 f}\right) \left(2-\frac{z_2}{2 f}\right) \frac{z_2}{2} + \left(2-\frac{z_1}{4 f}\right) \left(1-\frac{z_2}{2 f}\right) \frac{z_1}{4} \right] \left(1-\frac{z_1}{4 f}\right) \label{eq-B-expr} \\
C &=& -\frac{1}{f} \left[ \left(1-\frac{z_2}{2 f}\right) + \frac{\beta_1''}{\beta_2''} \left(1-\frac{z_1}{4 f}\right) \right] \left(1-\frac{z_1}{4 f}\right)
- \frac{1}{f} \left[ -\frac{\beta_2''}{\beta_1''} \frac{z_2}{2 f} \left(2-\frac{z_2}{2 f}\right) + \left(1-\frac{z_2}{2 f}\right) \left(1-\frac{z_1}{4 f}\right) \right] \label{eq-C-expr}
\end{eqnarray}
\end{widetext}
The algebraic complexity of writing out the expressions explicitly should not mask the simplicity of multiplying two-by-two matrices, usually numerically.
Note that the expression~(\ref{eq-D-expr}) for $D$ is algebraically identical to that for $A$~(\ref{eq-A-expr}), and it may be verified that $AD-BC = 1$. 

The $q$-parameter (we have dropped the $t$ subscript in this section for notational elegance) evolves according to the bilinear transformation law, and we require that the pulse repeat itself after propagation through one such ABCD matrix,
\begin{equation}
\frac{1}{q} = \frac{A + B/q}{C + D/q}
\end{equation}
which has the solution
\begin{equation}
\frac{1}{q} = \frac{D-A}{2 B} \pm i \frac{\sqrt{\displaystyle 1-\left(\frac{D+A}{2}\right)^2}}{B}
	\label{eq-q-eigen-equation}
\end{equation}

Since $D=A$ in our above analysis, we already see that $q$ is purely imaginary at $z=0$ i.e.~the pulse has zero chirp at the midplanes, as we would expect a breather to have.

At this stage, we can substitute numerical values for the various parameters (except $z_2$, which is what we seek) into
the expressions for the $A$, $B$, $C$ and $D$ elements~(\ref{eq-A-expr}--\ref{eq-D-expr}) and solve~(\ref{eq-q-eigen-equation}) numerically for $z_2$. While this is not difficult, and already yields a quick solution to the problem at hand, we can get further insight via a well-justified simplification as follows.

The $q$ parameter at the midplanes, where it is purely imaginary, is given by
\begin{equation}
\frac{1}{q_0} = \frac{2 |\beta_1''|}{\tau_0^2}
\end{equation}
where $\beta_1'' = 1.40 \times 10^{-27} \; \mathrm{s}^2/\mathrm{m}$ and input pulse width $\tau_0 = 5.67 \times 10^{-12} \; \mathrm{s}$.  Consequently, for such pulses, $1/q \approx 0$, and since $A=D$, this implies that $A=1$ in~(\ref{eq-q-eigen-equation}). 

With the notational substitutions
\begin{equation} 
x = \frac{z_2}{2 f}, \quad y=\frac{z_1}{4 f}, \quad \beta_r'' = \frac{\beta_1''}{\beta_2''}
\end{equation}
we get the necessary condition
\begin{equation}
\begin{array}{l}
\left[(1-y)^2 - (2-y) y\right] (1-x) - \beta_r'' y (2-y)(1-y) \\
\quad \displaystyle -\frac{1}{\beta_r''} (1-y)(2-x) x = 1.
\end{array}
\end{equation}
The solution of this equation is given by
\begin{equation}
x = \beta_r'' \frac{y(2-y)}{1-y}
\end{equation}
or, in terms of the initial variables,
\begin{equation}
z_2 = \left| \frac{\beta_1''}{\beta_2''} \right| \frac{\displaystyle \frac{z_1}{2}\left(2-\frac{z_1}{4 f}\right)}{\displaystyle \left(1-\frac{z_1}{4 f}\right)}
\end{equation}
which is the necessary condition in order to have a stable self-consistent Gaussian eigen-pulse (breather) solution to the dispersion-map problem. 

All that remains is for us to interpret the variables in terms of the original problem and numerically evaluate this expression to get the desired length $z_2$ of SMF-28 fiber in this dispersion map. The various numerical values are as follows:
\begin{eqnarray*}
\beta_1'' &=& 1.40 \times 10^{-27} \; \mathrm{s}^2/\mathrm{m}, \\
\beta_2'' &=& -2.12 \times 10^{-26} \; \mathrm{s}^2/\mathrm{m}, \\
\tau_0 &=& 5.67 \times 10^{-12} \; \mathrm{s}, \\
z_1 &=& 10^5\; \mathrm{m}
\end{eqnarray*}
Given the nature of the problem, we realize our time lens with the nonlinear fiber as described earlier~(\ref{eq-define-b}), so that
\begin{equation}
\frac{1}{f} = -4 \beta_{NL}'' \left(\frac{2 \pi}{\lambda} \right) \frac{n_2 I_p L_{NL}}{\tau^2}
\end{equation}
and take $\beta_{NL}''= \beta_2''$, $I_p = 3.62 \times 10^6\;\mathrm{W}/\mathrm{M}^2$ so that with fiber core area $A_{\mathrm{eff}} = 47\;\mu m$, we get $P = 8\; \mathrm{mW} = 9\; \mathrm{dBm}$. Also, we take $L_{NL} = z_2$ consistent with our choice of $\beta_{NL}''$.

The numerical solution (of the quadratic equation) for $z_2$ is equal to 7.00~km which is indeed the value ``approximately 7~km'' stated in the paper~\cite{art-Mu+-2000}. In spite of apparent exact agreement, we should be careful to appreciate that this analysis is a characterization of only the most important processes in this experiment. Possible sources for approximation include the fact that a DM soliton is only approximately Gaussian, and that we have represented the combined dispersive and nonlinear properties of the fiber segments by a single temporal lens. A better approximation may be to include several temporal lenses for each segment of fiber; this would make the algebraic expressions in this paper quite cumbersome to write down explicitly, but the numerical computation would not be much more difficult, since the matrices are only two-by-two in size, and comprise of purely real elements. The experimental configuration of~\cite{art-Mu+-2000} also includes several other elements which can affect the pulse shape, such as filters, fiber amplifiers and polarization controllers. 

\section{Hermite-Gaussian basis}
Our ABCD matrix formalism for pulse propagation applies to chirped Gaussian pulses. To analyze more complicated shapes, we can expand the given pulse shape in a basis of chirped Hermite-Gaussian functions, which form a complete orthonormal basis~\cite{bk-Haus,art-Lakoba-Kaup-1998}.  The Hermite-Gaussian function (we consider only unchirped Gaussians here for simplicity) of order $n$ is defined as the product of the Hermite polynomial of order $n$ with a Gaussian function,
\begin{equation}
\psi_n(t) \equiv H_n(t) \exp(-t^2/2),
\end{equation}
where, for example,
\begin{equation}
H_0 (t) = 1, \quad H_1(t) = 2 t, \quad H_2(t) = 4 t^2 -2.
\end{equation}

We can expand an arbitrary input amplitude $u_0(t)$ in this basis, analogous to expanding a field in terms of plane wave components, as in solution techniques of the standard parabolic diffraction equation by means of the Fourier transform, 
\begin{equation}
u_0(t) = \sum_{n=0} c_n H_n(t) \exp(-t^2/2)
\end{equation}
where because of orthogonality of the Hermite-Gaussians, the expansion coefficients are given by
\begin{equation}
c_n = \frac{1}{\sqrt{\pi}\, 2^n n!} \int_{-\infty}^{\infty} u_0(t) H_n(t) \exp(-t^2/2).
\end{equation}

The propagation equation~(\ref{eq-define-disperse}) defines the output pulse shape as the convolution of the input shape with a Gaussian kernel. Hermite-Gaussians, when convolved with a Gaussian, yield the product of a Hermite polynomial and a Gaussian~\cite{bk-Haus},
\begin{equation}
\begin{array}{lcr}
\multicolumn{2}{l}{\displaystyle \int_{-\infty}^{\infty} d t_0 \psi_n(t_0) \exp\left[ -\frac{a}{2} (t-t_0)^2\right] =} &\\ 
&\multicolumn{2}{r}{\displaystyle \sqrt{\frac{2\pi}{a+1}} {\left( \frac{a-1}{a+1}\right)}^{\frac{n}{2}}
\exp \left[ \frac{a\, t^2}{2 (a^2-1)}\right] \, \psi_n\left[\frac{a}{\sqrt{a^2-1}} t\right].}
\end{array}
\end{equation}
Taking as input the $n$-th Hermite-Gaussian mode $u_0=\psi_n(t)$ (which has width $\tau_0 = \sqrt{2}$), we evaluate the amplitude of this mode after propagation through distance $z$,
\begin{equation}
\begin{array}{lcr}
\multicolumn{2}{l}{
u_n(z, T) = \displaystyle \sqrt{\frac{1}{1+i \beta'' z}} {\left[-\frac{1+ \frac{i}{\beta'' z}}{1-\frac{i}{\beta'' z}} \right]}^{n/2}} &\\
&\multicolumn{2}{r}{\displaystyle \times \exp\left[ \frac{i}{2 \beta'' z} \frac{T^2}{1+\frac{1}{(\beta'' z)^2}}\right] \, \psi_n\left[ \frac{T}{\sqrt{1+(\beta'' z)^2}}\right]}
\end{array}
\end{equation}
which can be seen to agree with~(\ref{eq-width-chirp-propagate}).

A Hermite-Gaussian therefore maintains its shape during propagation, but adds a chirp (which is the same for all modes) and a scaling of the width according to~(\ref{eq-width-chirp-propagate}). Power conservation implies that the amplitude correspondingly scales down. The only term that is dependent on the order of the Hermite-Gaussian is a phase term; higher-order modes have greater phase advances, since their spectral content is higher. The important observation is that the orthogonality of the Hermite-Gaussian expansion is preserved, and so this expansion may be used to predict the pulse shape obtained by propagating an input pulse. Our formalism remains valid as long as the differential equation describing the propagation of a particular order Hermite-Gaussian is of the form~(\ref{eq-propagation-equation}), i.e.~the slowly-varying envelope approximation is valid. Therefore, we can expect that the lower-order expansions are usually valid; the results of applying our analysis to higher-order expansion terms generate the residual field corrections to the lower order results~\cite{art-TchofoDinda+-2000}. 

\section{Conclusions}
We have developed a $2\times 2$ {ABCD} matrix formalism for describing pulse propagation in media described by Maxwell's equations, accounting for dispersion, nonlinear and gain/loss mechanisms. The method is analogous to techniques used in CW beam diffraction analysis, and correspondingly similar phenomena can be predicted, such as chirp transformation, focusing, periodic pulse width expansion and narrowing etc. The spatial $q$ parameter has a time equivalent $q_T$ in accordance with the given space-time translation rules. The real and imaginary parts of $q_T^{-1}$ represent the chirp and the width of the pulse as a function of propagation distance $z$.

The propagation of various input pulse shapes can be described by expanding the given pulse in a basis of Hermite-Gaussian functions; the ABCD formalism applies to each Gaussian wave function separately. Propagation through a complicated system of optical elements is simple to calculate in terms of ABCD matrices: the resultant matrix is the cascaded product of the ABCD matrices of each of the individual elements with the appropriate ordering.  The overall $q_T$ parameter is given by a bilinear transformation in terms of the ABCD elements of the overall product matrix, exactly analogous to the spatial case. 

We have formulated ABCD matrices for pulse propagation in dispersive fibers, and for temporal lenses which can characterize self-phase modulation phenomena. A spatial dielectric interface translates to an interface between fiber segments of dissimilar GVD coefficient $\beta''$. The temporal equivalent of a curved dielectric interface is useful for characterizing the transition between such dissimilar $\beta''$ fibers with the added presence of fiber nonlinearities arising from the nonlinear index of refraction $n_2$. We have used these tools to characterize a reasonably complicated real-life system: calculation of the dispersion map for self-consistent stable propagation of a dispersion-managed soliton.

We believe this method of analysis forms a useful complement to conventional pulse propagation methods, such as the split-step Fourier transform numerical procedures~\cite{bk-Agrawal} which are substantially more computationally intensive. The ABCD approach is useful for clarifying the important dispersive and nonlinear focusing effects in dispersion-mapped  nonlinear fiber segments. Together with the variational approach~\cite{art-Anderson-1983}, based on modeling the pulse as a dynamical system characterized by a Hamiltonian functional~\cite{art-Muraki-1991}, the $q$-parameter offers an insight into pulse evolution from a theoretical standpoint.

\begin{acknowledgments}
This work was supported by the Office of Naval Research and the Air Force Office of Scientific Research.
\end{acknowledgments}

\end{document}